\documentclass[preprint,12pt,nopreprintline]{elsarticle}

\usepackage{amssymb}
\usepackage{amsmath}
\usepackage{algorithm}
\usepackage{algorithmic}
\usepackage{hyperref}
\usepackage{amsmath} 
\usepackage{tabularx}
\usepackage[expansion=false]{microtype}
\usepackage[english]{babel}
\usepackage[none]{hyphenat}
\usepackage{multirow}
\newcommand{\quotes}[1]{``#1''}
\usepackage{changepage}

\usepackage {hyperref}[colorlinks=true,
            linkcolor=blue,
            citecolor=blue,
            urlcolor=blue]

\begin{document}

\begin{frontmatter}



\title{Geolocation-Based User Clustering for Microservices: A Decoupled Gateway Approach to Hybrid Edge Orchestration}


\author{Anmol Chaudhary\corref{cor1}}
\cortext[cor1]{Corresponding author}
\ead{anmolc563@gmail.com}

\author{Dr. Rahul Mishra}
\ead{rmishra@niamt.ac.in}
\affiliation{organization={Department of Electronics and Computer Engineering},
            addressline={National Institute of Advanced Manufacturing Technology}, 
            city={Ranchi},
            country={India}}

\begin{abstract}

Modern low-latency applications face substantial challenges in satisfying the rigorous performance requirements of cloud-only architectures. The inherent propagation delays and centralized control dependencies are the reasons for the challenge. Despite the widespread adoption of edge computing in industrial environments to address latency issues, it introduces a range of new challenges, including fragmented control, distributed service management, and inconsistent performance under dynamic workloads and user mobility. The problem affects a range of industrial services, including real-time analytics, intelligent transportation, and location-enabled services, which require high service availability and predictable response times. To address these challenges, a novel two-layer orchestration architecture is proposed that simultaneously optimises service allocation and request routing. This model employs a decoupled control-plane paradigm that deviates from traditional centralized approaches. Rather than centralized gateways, the proposed paradigm employs distributed edge API gateways. These gateways employ a sliding window technique to optimise traffic routing by analysing data from peripheral gateways. Additionally, the proposed model employs DBSCAN clustering on geolocation data to create a demand-sensitive placement technique. The proposed model automatically adjusts the number of microservices in response to demand fluctuations. The exhaustive experimental study shows that this innovative design improves service availability and reduces end-to-end latency by up to 16.37\% compared to centralised systems, while ensuring latency variability remains within 2.13\% across multiple experimental runs. Ultimately, the comprehensive performance analysis demonstrates that decoupling data routing from orchestration can yield more efficient, scalable, and near-instantaneous data processing. 
\end{abstract}





\end{frontmatter}

\section{Introduction}\label{Introduction}

The shift from rigid monolithic architectures to loosely coupled microservice architectures marks a key evolution in software engineering, driven by the need for faster deployment cycles, greater flexibility, and elastic scalability in modern cloud-based applications. 

The packaging of application logic in traditional monolithic architectures impeded continuous integration and granular scalability, forcing developers to use a single executable file. By decomposing complete applications into multiple independent, lightweight services known as \quotes {microservices}, the microservices architecture addresses these limitations. These services are capable of autonomous development and deployment and can be created, deployed, and scaled independently. The microservices architectural style is well-suited to centralised cloud environments due to their almost unlimited resources, which enable elastic scalability, cost-effective resource use, and easier management \cite{Kaushik2025QoSMicroservices}. Accordingly, it is compatible with cloud computing. However, cloud-hosted microservices successfully addressed the computational scalability issue; however, they inadvertently introduced a critical impediment, namely the network \cite{shi2016edge}. As applications become more interactive and data-intensive, the physical distance between end users and centralised data centres becomes a defining limitation. 

The rapid expansion of Internet of Things applications and low-latency services has highlighted the intrinsic limits of the centralised cloud model, underscoring the need for new approaches. While centralized cloud computing provides considerable elastic scalability and resource pooling, its reliance on remote, centralized data centers results in unavoidable propagation delays limited by the speed of light. Network propagation delays make cloud-only execution inappropriate for real-time and interactive applications. Such constraints become evident in geographically extensive and diverse network regions, represented by countries like the Indian subcontinent, where end users may be separated from cloud data centers by thousands of kilometers. To address this, the cloud-edge continuum has recently emerged as a pivotal architectural transformation, enabling partial computation offloading to locations closer to end users, thereby decreasing latency and improving quality of service (QoS) \cite{Tang2024Industrial,Maia2024IntegratedCloudEdge,wang2017survey}.\\

Recent large-scale cloud service outages have further exacerbated the risks of relying solely on centralised cloud infrastructure, alongside latency concerns. Amazon Web Services (AWS), Microsoft Azure, and Google Cloud are the primary targets of service disruptions lasting up to 15 hours in 2025. These disruptions had a significant impact on major platforms such as Netflix, Snapchat, and various financial services \cite{aws2025pes,gcp2025status,azure2025status}. The potential for global disruptions to cascade from a single point of failure in a significant data center is underscored by these incidents, as shown in \cite{mao2017survey}. Cloud services may fail due to misconfigured control planes, issues with the core network, or service disruptions in a particular region, even if the application logic remains unaltered. These events emphasize distributed autonomy, as it enables peripheral services to continue serving requests independently during partial or complete disruptions, thereby improving the system's operational independence and resilience \cite{service_outage}. 

However, deploying computation at the edge introduces significant challenges related to resource constraints and operational cost \cite{zhou2023cost}.  It is economically unfeasible to maintain high-performance peripheral servers in regions with fluctuating or low user demand. Furthermore, numerous existing methodologies depend on intricate deep reinforcement learning (DRL)-based orchestration models \cite{wang2024gnn,chen2023balancing}. Although these models are robust, they are frequently cost-prohibitive to train, challenging to converge in dynamic environments, and challenging to comprehend and apply for inference.

Although these enhancements have been implemented, a significant issue with orchestrating hybrid environments remains. The majority of the extant research that has been conducted thus far has concentrated on the optimization of computation placement while disregarding the latency overhead that the control plane introduces. Additionally, although DRL-based methodologies and recent Large Language Model (LLM) approaches may be theoretically optimal, they frequently lack the lightweight agility and energy efficiency that peripheral nodes with restricted resources require. Specifically, the high inference costs of generative models render them impractical for battery-powered industrial controllers. The demand for an orchestration framework that simultaneously decouples the data plane to improve performance, ensures interpretability for operators, and maintains the cost-effectiveness of scale-to-zero models is profound.

The proposed work outlines a hybrid cloud-edge architecture (shown in Figure \ref{workflow_diagram}) that leverages a lightweight, interpretable, and heuristic-driven orchestration methodology to address these challenges. The proposed architecture utilizes a scale-to-zero methodology, whereby edge resources are dynamically created and dismantled according to spatio-temporal user density hotspots discovered using the DBSCAN algorithm \cite{nguyen2023serverless} \cite{zhang2021stdbscan}. Decoupling the control plane from the data plane is crucial. In contrast to conventional microservice architectures that experience delay due to accessing centralized service registries for each request, the proposed approach enables edge gateways to maintain a locally cached view of service instances and to make autonomous routing choices \cite{caprolu2024service}. Aggregated telemetry is communicated asynchronously with a centralized orchestrator, which performs deployment and removal choices independently of the request data plane. This facilitates low-latency decentralized request management while ensuring efficient and cost-effective resource use via a centralized control plane.

\subsection{Motivation \& Contributions}
The inherent latency problems of centralised cloud systems have become significant concerns as IoT applications evolve from passive data monitoring to decision-making and real-time action \cite{Li2025EnterpriseIoT}. Real-time applications, such as smart transportation and smart manufacturing, require instant responses, which centralised cloud architectures often can't deliver due to backhaul and network delays. At the same time, it's not always possible to keep edge systems online \cite{zhou2023cost, EnergyAware}. At the same time, it's not always possible to keep the edge system online. Edge resources aren't used during off-peak hours, leading to overprovisioning and higher running costs \cite{nguyen2023serverless}. As a result of these problems, there is a growing need for a scale-to-zero orchestration model that uses edge resources only when needed, not continuously.

The proposed work introduces a demand-aware framework for dynamic microservice orchestration at the edge. It uses geolocation-based user clustering and a decoupled gateway design to place services more efficiently and reduce end-to-end latency. The framework reduces backhaul traffic and speeds up the system by spreading gateway tasks across multiple gateways.

Also, the proposed model doesn't rely on complex AI models or specialised hardware accelerators. Such a design choice enables easier integration with existing cloud-edge infrastructure, reduces technical debt, and lowers capital expenditure (CAPEX). It also lowers technical debt and capital expenditure (CAPEX). Further, keeping the control plane separate from automated scale-to-zero processes makes it easier to follow Service Level Agreements (SLAs) for industrial applications that require rapid performance. In addition, this architectural separation reduces unnecessary operational costs and cloud data transfers. So, it makes the hybrid edge environment more sustainable and cost-effective.

The primary contributions of the proposed work are: 
\begin{itemize} 
\item \textbf {Lightweight Spatio-Temporal Orchestration:} We present an inevitable orchestration framework that uses real-time traffic monitoring and density-based clustering to automate service lifecycle management. Unlike computationally intensive learning-based models, the given approach offers a clear, low-overhead alternative for edge hardware.

\item \textbf{Resilient Hybrid Fallback Mechanism:} We present a dual-layer reliable model that dynamically deploys services at the edge for performance while keeping a cloud-based fallback that is always available. It ensures service continuity during demand spikes or when edge resources are unavailable.

\item \textbf{Decoupled Gateway Architecture:} 
We present a distributed API gateway architecture that separates request routing from centralized service discovery by keeping a cached view of the service locally. The given method reduces the control-plane dependency, improving end-to-end latency and increasing the system's fault tolerance to network failures.

\end{itemize}

\subsection{{Roadmap of the paper}}
The structure of the paper is organized as follows: Section 2 discusses the state-of-the-art orchestration strategies with a brief discussion on trade-offs among heuristic-based, bio-inspired, and learning-based strategies. Section 3 presents the proposed spatio-temporal orchestration framework, including a decoupled gateway architecture, an optimization formulation, and a DBSCAN-based demand-sensitive scaling mechanism. Section 4 presents the evaluation of the proposed orchestration scheme with a focus on latency reduction, scalability, and tail latency against centralized baselines. Section 5 discusses the potential industrial effects of the proposed method with operation/edge infrastructures, while Section 6 presents the conclusion of the proposed method along with possible future research directions.

\section{Literature Survey}\label{literature_survey}

The process of optimizing hybrid cloud-edge architectures has evolved into a critical research field, emphasizing lightweight methods to minimize latency and energy usage. The current section reviews contemporary contributions in three main areas: heuristic-based microservice placement, container cold-start mitigation, and decentralized service discovery protocols.

\subsection{Heuristic vs. Learning-Based Orchestration}
Efficient service placement is a critical, often NP-hard, challenge in the edge computing paradigm that has a direct impact on system performance, such as latency, resource utilization, and receptivity. In view of the latency and resource constraints of edge gateways, optimization approaches should aim for bounded complexity and deterministic execution time over asymptotic optimality.



Similarly, for algorithmic efficiency, Panadero et al. \cite{Tigo} proposed a two-stage heuristic optimization method, where an initial preliminary candidate selection is followed by a precise nonlinear optimization approach to solve the service placement problem in edge computing, achieving lower execution time compared to traditional genetic-algorithm-based solutions.

Kazmi et al. \cite{santos2024dynamic} performed a comparative analysis of nature-inspired algorithms for dynamic service placement and concluded that, although swarm intelligence models have high accuracy, they often have slower convergence speeds than greedy algorithms when handling rapid user mobility.


Unlike these lightweight strategies, recent studies have examined several deep learning models. Chen et al. \cite{wang2024gnn} introduced a deep reinforcement learning model assisted by graph neural networks. The approach greatly augmented the load balancing in complex networks. However, this approach has a high computational cost for training, making it less efficient for resource-limited edge gateways.

More recently, AI-driven methods for edge orchestration have gained recognition due to their ability to deal with complex system dynamics. Peng et al. \cite{AIservices} proposed a precise hybrid orchestration framework using the Action-Masked Deep Q-Networks (AMDQN) technique, which achieved up to a 41.8\% and 10.13\% reduction in latency delay in service chaining compared to greedy baselines and when compared with advanced algorithms. 

Similarly, some work on generative AI-based orchestration has been conducted. Feng et al. \cite{GenAI} concluded that Large Language Models (LLMs) can be utilized to guide microservice deployment decisions during traffic surges, achieving latency improvements of approximately 22.8\% .

However, such learning-intensive approaches introduce practical challenges in real-time edge environments. AMDQN-based frameworks typically rely on iterative training processes and extensive state exploration, while LLM-driven solutions incur substantial inference latency and memory overhead \cite{Yu2026Energy}. Even though these are very good at optimizing, their complicated operation and unpredictable performance can make them less useful for edge gateways that need to make quick decisions in milliseconds and have limited resources.

\subsection{Mitigating Container Cold-Start Latency}
As edge environments require rapid scaling, it is essential to mitigate container startup time. Buitrago et al. \cite{wang2021parsec} proposed an adaptive serverless platform that has the capability to overcome the cold-start problem by using a \quotes {snapshot} technique to pre-initialize the containers and thus achieve a reduction in the startup time of around 40\%.

For the container reuse issue, Zijun Li et al. \cite{wang2022hotc} introduced a framework for managing container runtimes that utilizes an exponential smoothing algorithm to predict the reuse of the containers in order to maintain a set of \quotes {warm} containers to eliminate the initialization delay for frequent requests.

More recently, architectural innovations have surfaced. Galvankar et al. \cite{galvankar2025alphaboot} proposed a hardware-accelerated solution using SmartNICs to cache container images at the network interface level, effectively bypassing the OS storage stack to speed up image retrieval and deployment. 

Similarly, Zhiqing Tang et al. \cite{Tang_2024} designed a layer-aware scheduler for Docker containers to optimize the placement of containers by analyzing the shared image layers, ensuring that nodes with cached base layers are given priority to mitigate download costs.

Although these approaches are effective, they often overlook the coupling issues. Luqi Sun et al. \cite{zhang2025coldstart} addressed the problem by designing a nested container architecture that shares runtime environments; however, their model, while fast, inadvertently increases microservice coupling, complicating updates.

\subsection{Decentralized Discovery and Spatial-Temporal Clustering}
Efficient decentralized discovery directly impacts orchestration latency by reducing control-plane round trips, making it a first-order concern rather than a supporting optimization.

To reduce control-plane dependency, research is shifting toward decentralized architectures. Derya Birant et al. \cite{zhang2021stdbscan} used Spatial-Temporal DBSCAN to allocate users, showing that density-based clustering outperforms fixed-grid clustering methods for irregular user hotspots in real time.

To complement this, K. Horvath et al. \cite{horvath2025location} designed an efficient location-based service discovery protocol for IoT and edge computing in the 6G era, which used geospatial coordinates in DNS responses using Location Resource Records and allowed clients to connect to nearby edge services based on their location, reducing lookup latency and centralized registry dependencies.

Regarding architectural decoupling, Ren et al. \cite{ai2022triring} developed an asynchronous service provisioning framework that decouples the request dispatching plane from the microservice orchestration plane, enabling edge nodes to process requests independently while eventually synchronizing state.

Nevertheless, the current centralized decision-making model, such as the network bandwidth allocation model proposed by Mao et al. \cite{li2024resource}, tends to rely on a global network view, which may become a performance bottleneck during network partitions.\\

\begin{table*}[t]  
\begin{adjustwidth}{-1cm}{-1cm}
\centering
\scriptsize
\setlength{\tabcolsep}{4pt}
\renewcommand{\arraystretch}{1.3}
\newcolumntype{Y}{>{\centering\arraybackslash}X}
\begin{tabularx}{1.27\textwidth}{|p{1.5cm}|Y|Y|Y|Y|Y|Y|Y|}
\hline
\multirow{2}{*}{\textbf{Schemes}}
& \multicolumn{7}{c|}{\textbf{Comparison parameters}} \\
\cline{2-8}
& \textbf{Focus}
& \textbf{Orchestration Logic}
& \textbf{Low Comp. Overhead}
& \textbf{Scale-to-Zero}
& \textbf{Decoupled Gateway}
& \textbf{Spatio-Temp. Cluster}
& \textbf{Hybrid Fallback} \\
\hline

Peng et al. \cite{AIservices}
& Hybrid Orchestration
& Hybrid (Rule-based + AI-assisted)
& \texttimes
& \texttimes
& \checkmark
& \texttimes
& \checkmark \\
\hline

Feng et al. \cite{GenAI}
& Service Placement / Optimization
& LLM-based Reasoning (RAG + CoT)
& \texttimes
& \texttimes
& \texttimes
& \checkmark
& \texttimes \\
\hline

Kazmi et al. \cite{santos2024dynamic}
& Service Placement
& Nature-Inspired Algo
& \texttimes
& \texttimes
& \texttimes
& \texttimes
& \texttimes \\
\hline

Birant et al. \cite{zhang2021stdbscan}
& User Density Allocation
& Density-Based Clustering
& \checkmark
& \texttimes
& \texttimes
& \checkmark
& \texttimes \\
\hline

Horvath et al. \cite{horvath2025location}
& Service Discovery
& Geospatial Embedding
& \checkmark
& \texttimes
& \checkmark
& \texttimes
& \texttimes \\
\hline

Ren et al. \cite{ai2022triring}
& Async Provisioning
& Asynchron. Queue
& \checkmark
& \texttimes
& \checkmark
& \texttimes
& \texttimes \\
\hline

Buitrago et al. \cite{wang2021parsec}
& Serverless Latency
& Snapshot and Restore
& \checkmark
& \checkmark
& \texttimes
& \texttimes
& \texttimes \\
\hline

Chen et al. \cite{wang2024gnn}
& Load Balancing
& Deep RL + GNN
& \texttimes
& \texttimes
& \texttimes
& \checkmark
& \texttimes \\
\hline

\textbf{Proposed Work}
& \textbf{Hybrid Orchestration}
& \textbf{Heuristic + DBSCAN}
& \textbf{\checkmark}
& \textbf{\checkmark}
& \textbf{\checkmark}
& \textbf{\checkmark}
& \textbf{\checkmark} \\
\hline

\end{tabularx}

\caption{Feature comparison of existing state-of-the-works  and the proposed method}
\label{tab:comparison1}
\end{adjustwidth}
\end{table*}

Table \ref{tab:comparison1} provides a qualitative comparison between our proposed framework and existing solutions. To the best of our knowledge, no prior work jointly addresses lightweight orchestration, decentralized gateway-level decision-making, and hybrid cloud–edge fallback guarantees within a single architectural framework.

\section{Proposed System Architecture and Methodology}

The given section details the Spatio-Temporal Orchestration Framework, a hybrid edge-cloud architecture designed to minimize latency through a strict decoupling of the control and data planes. The system distinguishes between Distributed Edge Gateways, which perform autonomous low-latency routing, and a Centralised Cloud Core, which optimises global placement decisions asynchronously.

\subsection{System Architecture Overview}
 We model the infrastructure as a set consisting of a centralized cloud core $\mathcal{C}$ and a distributed set of edge nodes $\mathcal{N} = \{e_1, e_2, ..., e_n\}$. User devices constitute the dynamic user plane U(t), generating geo-tagged requests ${r_i}$ with coordinates $(\phi_u, \lambda_u)$.


\begin{figure}[htbp]
\centering
\includegraphics[width=\linewidth]{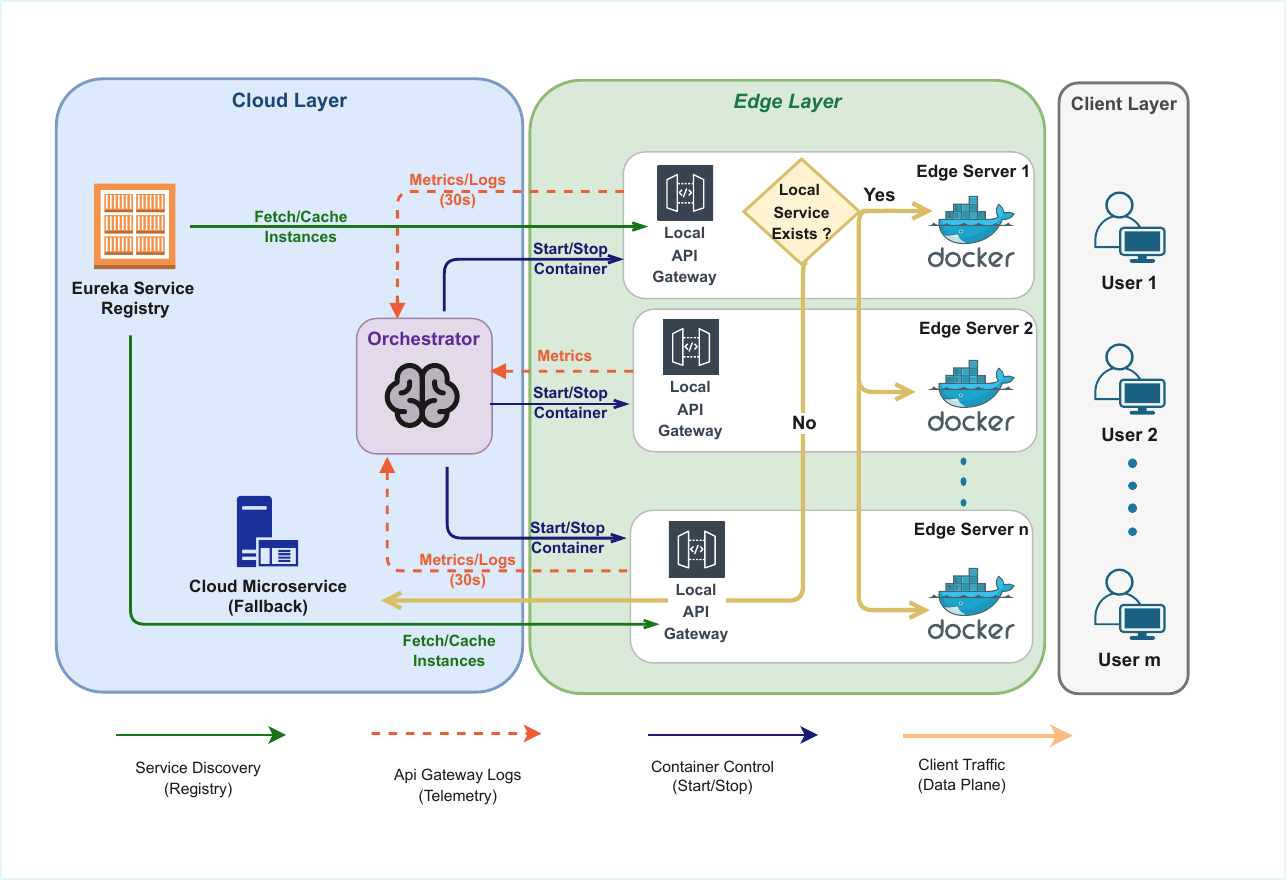}
\caption{Operational Workflow of the Proposed Hybrid Edge–Cloud Architecture}
\label{workflow_diagram}
\end{figure}

As shown in Figure \ref{workflow_diagram}, the proposed architecture preserves a strictly decoupled control-data plane topology and depends on an eventual consistency model. The Data Plane (Edge Layer) focuses on immediate availability, responding to requests from locally cached registry entries, thereby decoupling the critical path from WAN latency. Simultaneously, the Control Plane (Cloud Layer) maintains consistency through asynchronous synchronization loops, where decisions for the service instantiation and removal related to the orchestration of aggregated telemetry never block real-time execution. 

Unlike prior edge orchestration frameworks that rely on synchronous control-plane involvement, our architecture enforces a strictly non-blocking data plane, guaranteeing bounded request latency even under control-plane failures. \\

To achieve a clear separation of concerns, the proposed framework is structured into three logical layers as shown in Figure \ref{workflow_diagram}: Cloud Core, Edge Layer, and Client Layer. Each layer is responsible for a specific aspect of orchestration, execution, and request generation.

\subsubsection{Cloud Core}
The cloud tier serves as the central control layer of the proposed architecture and is responsible for maintaining a consistent global system state, coordinating logic, and executing fallback services. It comprises three key components:
\begin{itemize}
    \item \textbf{Central Orchestrator:} The Central Orchestrator is the system's decision-making engine. Instead of directly processing user traffic, it uses telemetry logs from edge nodes, maintains a 1-minute sliding window ($W_t = 60 s$), and executes decision logic to trigger the Deploy and Kill container lifecycle events.
    \item \textbf{Service Registry (Eureka Server):} The global service registry \(R_c\) maintains the instances of active microservices.
\[R_c(t)=\{S_1, S_2, \dots, S_M\}\] 

It dynamically adds and removes Edge instances as they are instantiated and registers the Cloud Fallback instance as a permanent entry.
\item \textbf{Cloud fallback Microservice:} The fallback microservice operates as the highly available microservice that maintains the uninterrupted service availability when the orchestrator has not deployed any active edge instance or Edge Nodes are scaled to zero. It also ensures fault tolerance and avoids hard downtime.
\end{itemize}

\subsubsection{ Edge Layer}
The dynamic execution environment is hosted on nodes strategically positioned at the network edge layer. Each Node \( e_i\) hosts two components: 

\begin{enumerate}
\item \textbf{Smart Edge Gateway:} The gateway acts as the local traffic manager at the edge, deciding where each user request should go based on the available service instances. And at each edge node, it fulfills three responsibilities:

\begin{itemize}

    \item \textbf{Local Caching-} Each gateway maintains a local copy \(\hat{R_c}(t)\) of the cloud registry. Also, the periodic synchronization removes cloud lookup latency from the request path.
    
    \item \textbf{Telemetry Aggregation-}   The edge gateway collects average latency metrics ($L_{avg}$), request counts ($r_{count}$), and user request coordinates $r_i(\phi_u, \lambda_u)$. To reduce network bandwidth use, they are compressed and sent to the Orchestrator in asynchronous batches every 30 seconds ($t_{log}$).
    
    \item \textbf{Hybrid Routing-} The routing ensures ultra-low-latency routing even when cloud RTT is high. The routing logic directs requests to the local instance if available in the cache; otherwise, it redirects to the cloud fallback.
   
    \[
    \text{Route}(r_i) =
    \begin{cases}
    \text{$S_{local}$}, & \text{if } e_i \in \hat{{R_c}}(t) \\
    \text{$S_{cloud}$}, & \text{otherwise}
    \end{cases}\]

\end{itemize}

\item \textbf{Dynamic Microservice Container:} These are temporary instances of the application. Only when the user density exceeds the activation threshold ($T_{high}$) are they deployed, and they are removed when the demand falls below the hysteresis limit ($T_{low}$).
\end{enumerate}

\subsubsection{ Client Layer} The end-user devices, like mobile phones and IoT sensors that produce the traffic, constitute the bottom layer. Customers communicate only with the closest Edge Gateway via an API endpoint and are oblivious to the hybrid topology. Significantly, the control-plane operations such as clustering, deployment decisions, and registry synchronization are executed asynchronously and never lie on the critical request path, preserving low-latency guarantees at the edge.\\

Unlike existing edge–cloud orchestration frameworks that either rely on synchronous cloud control, centralized service discovery, or learning-based placement policies, the proposed framework introduces a strictly decoupled control–data plane design. The data plane operates autonomously at the edge using cached service state and local routing decisions, guaranteeing bounded request latency even during cloud unavailability. Meanwhile, the control plane performs spatial–temporal optimization asynchronously, combining density-based clustering with hysteresis-driven deployment to avoid oscillatory behavior. The inclusion of a permanent cloud fallback microservice further ensures service continuity during scale-to-zero edge operations, which prior architectures do not jointly address.

\subsection{System Architecture and Problem Statement}
In this section, we present an interesting high-level system model and orchestration logic that works with any software platform or deployment tool. It clearly explains the roles of important architectural parts like edge gateways, service registries, and orchestrators, focusing on how they help reduce delay and improve resource management. Comprehensive insights into the real deployment environment, tools, and technologies are provided in the implementation subsection.\\
The main operational problem is to mitigate system latency ($L$) while augmenting resource utilization ($U$) without adding control-plane overhead to the request path.

\begin{enumerate}
    \item \textbf{Optimization Objective:} At each separate 
    decision time step $t_k$, the Orchestrator chooses a deployment configuration $D(t_k)$, which is a subset of the set of all available edge nodes $\mathcal{N}$ (i.e., $D(t_k) \subseteq \mathcal{N}$). The objective is to minimize a weighted cost function $J(D)$ that balances system latency, operational costs, and service reliability. The optimization problem is set out like this:

\begin{equation}
\min_{D(t_k)} J(D(t_k)) = \alpha \cdot \mathbb{E}[L(D(t_k))] + \beta \cdot C(D) + \gamma \cdot P_{\text{unavail}}(D)
\label{eq:optimization}
\end{equation}

    And the operational cost term $C(D)$ is defined as the total resource cost included by active edge deployments:

    \begin{equation}
           C(D) = \sum_{e_i \in D} (c_{\text{idle}} + y_i \cdot c_{\text{run}})
           \end{equation}

where $y_i \in \{0,1\}$ indicates if the node is actively executing a container.\\
 
 The framework inherently promotes scale-to-zero behaviour in low-demand areas since, $c_{\text{idle}}$ is much lower than $c_{\text{run}}$.

    Subject to the physical resource constraint, the minimization is bounded by the capacity of the edge nodes:\\
    
    Let $\mathcal{S} = \{S_1, S_2, \dots, S_M\}$ denote the set of candidate microservices considered by the orchestrator for deployment decisions, where $M$ is the total number of such services.

    $$\sum_{k=1}^{M} \text{Res}(S_k) \cdot x_{i,k} \leq \text{Cap}(e_i), \quad \forall e_i \in D(t_k)$$Where $\text{Cap}(e_i)$ represents the CPU/RAM limits of the edge node, ensuring deployments do not exceed hardware capacity. And $x_{i,k} \in \{0, 1\}$ is a binary decision variable indicating whether microservice $S_k$ is deployed on edge node $e_i$.

    where the components are defined as:
    \begin{itemize}
        \item $c_{\text{idle}}$: The basic cost of keeping an edge node idle.
        \item $c_{\text{run}}$: The extra cost of running an active microservice container. 
        \item $\mathbb{E}[L(D)]$: The predicted end-to-end latency for the configuration that is chosen.
        \item $C(D)$: The computing power cost of executing microservices on the chosen edge nodes.
        \item $P_{\text{unavail}}(D)$: A penalty function that is required for measuring the risk of service unavailability.
        \item $\alpha, \beta, \gamma$: Non-negative weighing coefficients that prioritize Quality of Service (QoS), cost-effectiveness, and reliability first.
        \item $M$: The total number of microservices that can potentially be deployed on an edge node.
        \item $S_k$: The k-th microservice in the system.
        \item $Res(S_k)$: The resource requirement of microservice i.e., CPU cores, memory, or a weighted combination.

    \end{itemize}

\item \textbf{Latency Optimization through Cached Routing via Control–Data Plane Decoupling:}

 In typical cloud-centric microservice designs, the control plane is closely linked to request routing and service discovery \cite{silva2021discovery,yang2021replaceable}. It adds an extra delay to every request sent to the cloud. The total latency can be written as: \[L_{\text{traditional}} = RTT_{\text{cloud}} + RTT_{\text{registry}} + T_{\text{processing}}\]

On the other hand, the suggested architecture separates the control plane from the data plane by moving all service discovery and routing decisions to the edge gateway. The request latency becomes: \[L_{\text{proposed}} = RTT_{\text{edge}} + T_{\text{processing}}\] since registry lookups happen locally from cached state.

This architectural decoupling gradually removes cloud round-trip delays from the key request channel, enabling consistently low-latency execution even when workloads are spread across different locations.
\\

In a typical microservice architecture, a gateway queries the Service Registry for every request, which incurs a Round Trip Time ($RTT_{registry}$) penalty.
We suggest a Cached-Registry Model. At time $t$, let the $Cache_e(t)$ be the local set of service instances that Edge Gateway $e$ has saved. The routing function $Route(r_i)$ is defined as:$$Route(r_i) = \begin{cases} \text{Localhost} & \text{if } \exists \text{ Instance} \in Cache_e(t) \\ \text{Cloud}_{fallback} & \text{if } Cache_e(t) = \emptyset \end{cases}$$The $RTT_{registry}$ becomes closer to zero because the lookup happens in local memory. The overall request latency, $L_{r_i}$, is now: $$L_{r_i} \approx T_{network} + T_{processing}$$

 \item \textbf{Computational Complexity Analysis:} To evaluate the theoretical scalability of the framework, we analyze the decision complexity. The Data Plane Routing utilizes a hash-based cache lookup, which has been theoretically proven to achieve $O(1)$ amortized time complexity for key-value retrieval \cite{Kirsch2010HashBasedTF}\cite{carter1979universal}. It ensures constant-time access for service discovery regardless of the total number of registered microservices. The Control Plane Clustering, i.e., DBSCAN operates with $O(n \log n)$ complexity using spatial indexing, which remains computationally feasible for the asynchronous control loop \cite{nlogn}.

\item \textbf{Spatial Clustering With DBSCAN:} User demand clusters are discovered with DBSCAN by grouping dense areas of users without making any assumptions about fixed grid partitions \cite{zhang2021stdbscan}\cite{starczewski2021dbscan,10433132}. 

Unlike centroid-based algorithms like K-Means \cite{macqueen1967some}, which assume spherical clusters and require a predefined number of centroids ($k$), DBSCAN is uniquely suited for this problem for three reasons: 

\textbf{(1) Arbitrary Cluster Shapes:} User mobility often follows roads or coastlines, forming non-convex shapes that K-Means fails to capture.

\textbf{(2) Noise Robustness:} DBSCAN explicitly classifies low-density outliers as \quotes {noise}, allowing our framework to route isolated users to the Cloud Fallback and prevent inefficient edge deployment. 

\textbf{(3) Dynamic Discovery:} It does not require knowing the number of hotspots ($k$) a priori, allowing the system to adapt dynamically to changing crowd patterns.\\

Earlier, a user $u$ is considered part of a cluster if the density of their local neighbourhood satisfies the condition: $$|N_{\epsilon}(u)| \geq \text{MinPts}$$ where $N_{\epsilon}(u)$ denotes the $\epsilon$-neighborhood centered at $u$.
    The radius $\epsilon$ represents a real-world physical distance that is determined using the Haversine metric to account for geospatial adjacency for a user at $(\phi_u, \lambda_u)$ and an edge node at $(\phi_e, \lambda_e)$. Users identified as noise are routed to the cloud fallback service, ensuring efficient hybrid execution.
    \begin{equation}
    \small d = 2R\,\operatorname{arcsin}\!\left(
    \sqrt{sin^{2}\!\left(\tfrac{\Delta\phi}{2}\right)+ \cos\phi_{u}\cos\phi_{e}\,\sin^{2}\!\left(\tfrac{\Delta\lambda}{2}\right)
    }
    \right)
    \end{equation}
    $$T_{prop} \approx \frac{d}{c_{fiber}} \cdot \tau_{routing}$$

    where, $R$ is the Earth's radius, and $(\phi_u,\lambda_u)$ and $(\phi_e,\lambda_e)$ are the latitude and longitude of the user and the edge server, respectively. The differences between them are shown by $\Delta\phi$ and $\Delta\lambda$. 

    The Haversine great-circle distance is $d$, the propagation speed in optical fiber is $c_{\text{fiber}}$, the routing overhead factor is $\tau_{routing}$, and the predicted propagation delay is $T_{\text{prop}}$.\\

    At time $t$, DBSCAN found the collection of spatial demand hotspots $\mathcal{H}_t = \{h_1, h_2, ..., h_m\}$. Each hotspot is a group of user requests that are close together. If an edge node $e_i$ is within the spatial influence of an active hotspot, it is chosen for service deployment:

    \[
    e_i \in D(t_k) \iff \exists h_j \in \mathcal{H}_t \text{ such that } d(e_i, h_j) \leq \epsilon
    \]
    
    The above condition directly links clusters of spatial demand to deployment decisions, making sure that microservices are only created in areas where there is a lot of user activity.

    \item \textbf{Temporal Orchestration with Sliding Window:} The Orchestrator collects metrics over a sliding time window of length $W$ to reduce the chance of oscillatory deployments, which are often called container chattering. Let $\hat{\lambda}_t$ be the smoothed rate at which requests come in at time $t$. The deployment state $S_e(t+1)$ for an edge node $e$ is changed according to the following hysteresis function:

     \begin{equation}
    S_e(t+1) = 
    \begin{cases} 
        1, & \text{if } \hat{\lambda}_t \geq T_{\text{high}} \quad (\text{Deploy}) \\
        0, & \text{if } \hat{\lambda}_t < T_{\text{low}} \quad (\text{Terminate}) \\
        S_e(t), & \text{otherwise} \quad (\text{Maintain State})
    \end{cases}
 \end{equation}

where $S_e(t) \in \{0,1\}$.
\\
The hysteresis constraint $T_{\text{high}} > T_{\text{low}}$ keeps the system stable by stopping deployment behavior from oscillating when traffic is temporary or bursty. This makes sure that containers are deployed and stopped only as demand changes over time, which keeps the control plane from thrashing.

\end{enumerate}

The proposed framework prioritizes deterministic latency guarantees over optimal but computationally intensive orchestration strategies. Spatial clustering enables demand-aware deployment without global synchronization, while temporal hysteresis stabilizes container lifecycle decisions under bursty workloads. Cached service discovery removes registry round-trip delays from the critical path, and the cloud fallback mechanism ensures availability without forcing persistent edge deployments.

\subsection{System Implementation}
The section describes the practical realization of the proposed hybrid edge–cloud architecture with the objective of validating the architectural and methodological concepts introduced earlier. The implementation emphasizes functional correctness, architectural feasibility, and control–data plane separation rather than system-specific performance optimization. Widely adopted, production-grade technologies are used to demonstrate that the proposed design can be deployed on heterogeneous edge infrastructure without relying on specialized hardware or proprietary platforms. 

The implementation closely follows the theoretical model from the previous section, with a direct link between software modules and analytical components. The Edge Gateway handles the routing logic for the data plane, the Cloud Orchestrator handles the optimization and hysteresis policy for the control plane, and the service registry makes sure that deployment decisions and request routing are always consistent. All control-plane interactions are asynchronous and do not form part of the critical request path.\\

Algorithm \ref{Algorithm} presents the integrated execution workflow of the proposed hybrid edge–cloud framework. 
The workflow is organized into three logical phases: 
\begin{enumerate}
    \item Real-time request routing executed entirely at the edge data plane.
    \item Asynchronous telemetry aggregation from distributed edge gateways. 
    \item Spatial–temporal orchestration decisions performed at the cloud control plane.
\end{enumerate} 

Control-plane operations are strictly decoupled from the request execution path and influence only future routing and deployment decisions, ensuring bounded request latency at the edge while enabling adaptive resource orchestration.

\begin{algorithm}
\caption{Integrated Hybrid Edge-Cloud Execution Workflow}
\begin{algorithmic}[1]
\REQUIRE Incoming Request Stream $Q$, Thresholds $\tau_{high}, \tau_{low}$
\ENSURE Optimized Request Routing and Resource Allocation

\STATE \textbf{Phase I: Request Routing (Data Plane)}
\FORALL{Incoming Request $r_i(\phi_{u}, \lambda_{u}) \in Q$ }
    \STATE \COMMENT{Initial handshake}
    \STATE $IP_{edge} \leftarrow \text{Orchestrator}.\text{GetNearestEdge}(r)$
    \STATE Request $r_i$ forwarded to Edge Gateway (Node $e_i$ with $IP_{edge}$)
    \STATE ${Cache_e} \leftarrow \text{SyncRegistryFromEureka}()$
    \STATE \COMMENT{Check for local service availability}
    
    \IF{$\exists S_{local} \in {Cache_e}$ on Node $e_i$} 
    
        \STATE Route $r_i \to S_{local}$ \COMMENT{Lowest Latency path}
        \STATE $\text{Latency} \leftarrow \text{Low}$
    \ELSE
        \STATE Route $r_i \to S_{cloud}$ \COMMENT{Cloud Fallback High Latency Path}
        \STATE $\text{Latency} \leftarrow \text{High}$
    \ENDIF
    
    \STATE \textbf{Log Generation:}
    \STATE $\mathcal{L}.\text{append}(\{e_i, \text{Timestamp}, \text{Latency}\})$
\ENDFOR

\STATE \textbf{Phase II: Metric Aggregation}

\IF{Reporting Interval ($T_{report}$) Elapsed}
    \STATE Transmit $\mathcal{L}$ from $G_{edge} \to \mathcal{O}$ (Cloud Orchestrator)
    \STATE $\mathcal{O}.\text{UpdateSlidingWindow}(\mathcal{L})$
\ENDIF

\STATE \textbf{Phase III: Orchestration Decision (Control Plane)}
\STATE $\hat\lambda_t \leftarrow \text{CalculateLoadDensity}(\mathcal{O})$

\IF{$\hat\lambda_{t} > \tau_{high}$ \AND No Local Instance on $e_i$}
    \STATE \textbf{Action: Scale Out (Edge)}
    \STATE $\text{DockerRun}(e_i, \text{ServiceImage})$
    \STATE Register new $S_{local}$ with Eureka
    \STATE \textit{Outcome: Future requests follow Low Latency Path}

\ELSIF{$\hat\lambda_{t} < \tau_{low}$ \AND Local Instance Exists on $e_i$}
    \STATE \textbf{Action: Scale In}
    \STATE $\text{DockerStop}(e_i, \text{ServiceInstance})$
    \STATE Deregister $S_{local}$ from Eureka
    \STATE \textit{Outcome: Future requests follow Fallback Path}

\ELSE
    \STATE \textbf{Action: Stabilize}
    \STATE Maintain current topology
\ENDIF

\end{algorithmic}
\label{Algorithm}
\end{algorithm}

\begin{figure}[htbp]
\centering
\includegraphics[width=\linewidth]{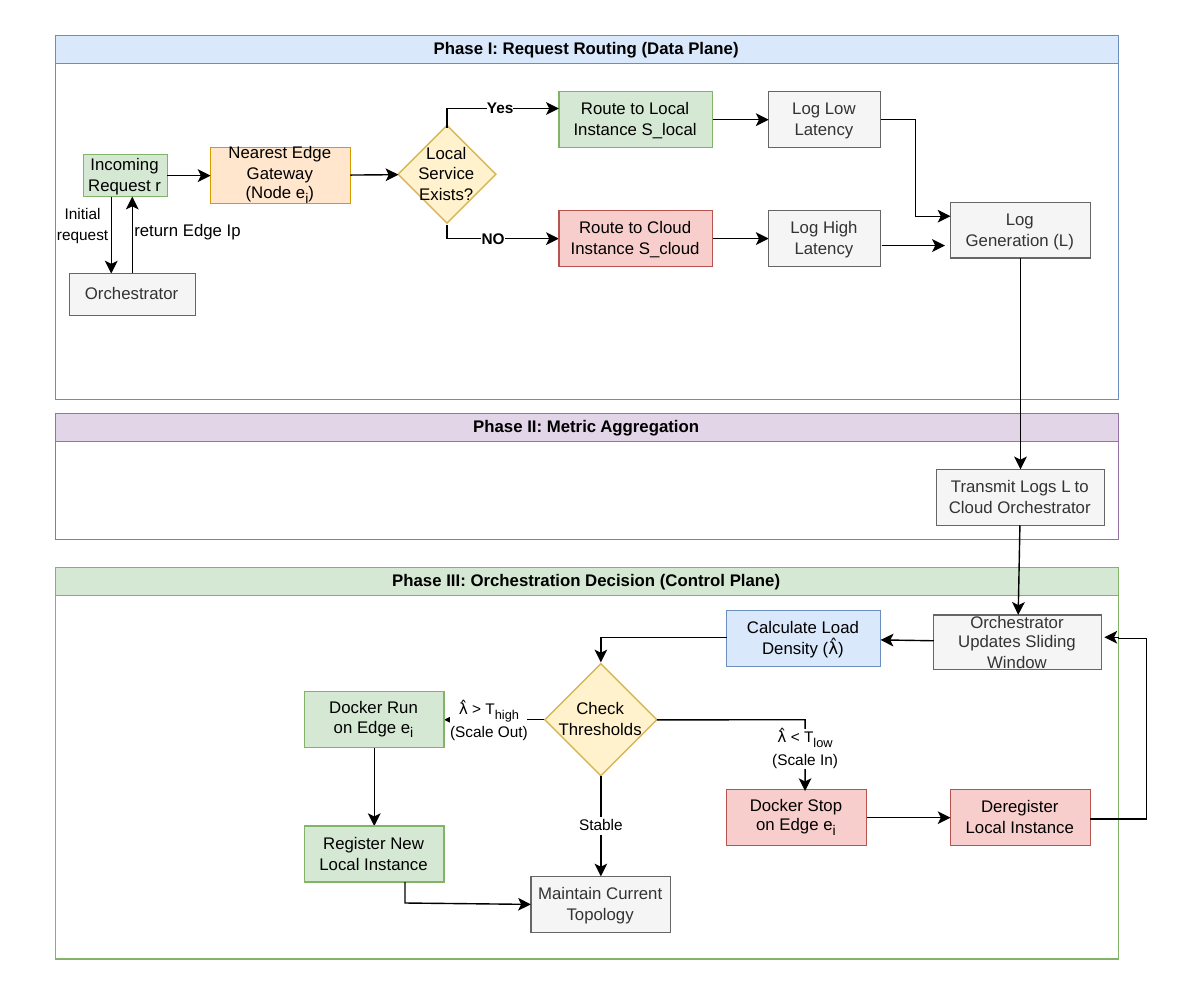}
\caption{Integrated Algorithmic Flow for Hybrid Service Execution and Dynamic Orchestration.}
\label{psudoCode}
\end{figure}

Figure \ref{psudoCode} visualizes a flowchart that defines the integrated logic that controls the system in three different phases. Phase I demonstrates the real-time routing decisions made at the network edge. Phases II and III emphasize the asynchronous control loops and define how aggregated metrics make the cloud orchestrator evaluate demand against thresholds. And finally, it decides when Docker containers are deployed or terminated to make the best use of resources.

The following subsection describes the implementation details of the proposed system, including the technological stack, the distributed gateway design, the telemetry pipeline, and the orchestration logic that will enable the theoretical model to operate.

\begin{enumerate}
    \item \textbf{Technology Stack and Environment:}
    The selected technology stack directly reflects the architectural components defined in the system model and ensures a one-to-one mapping between analytical entities and deployed software modules.
    \begin{itemize}
        \item Microservices Framework: Java Spring Boot (v3.x) for creating lightweight, independent services.
        \item Service Discovery: Netflix Eureka is used for the global service registry.
        \item  Edge Gateway: For the distributed edge gateways, customized Spring Cloud Gateways with local caching and routing filters are used.
        \item Container Runtime: Docker Engine is used, which also enables \quotes {scale-to-zero} capability on edge nodes \cite{haruna2022docker}.
        \item Orchestrator: For Orchestrator, a custom Python-based control module utilizing the Docker Remote API for lifecycle management is used.
    
    \end{itemize}

\item \textbf{The Distributed Gateway Implementation (Data Plane):}
    The Edge Gateway is the critical component responsible for the Fast Path of traffic. In traditional architectures, gateways are only simple proxies while we introduced it as a Smart Caching Gateway, capable of autonomous routing. To achieve this, the implementation integrates two distinct mechanisms:

        \begin{enumerate}
        \item Registry Caching Plan: To minimize the latency, the Edge Gateway doesn't query the Cloud Eureka server for every incoming request. Instead, it maintains a local cache of the service registry. Every 30 seconds, 
        a background synchronization thread invokes a registry refresh operation, which abstracts the native registry fetch mechanism to update the local cache, which decouples the availability of the edge from the cloud connectivity. If the backhaul link goes down, the gateway's cached registry keeps sending traffic to local instances.

        \item Latency-Aware Routing Logic: The routing logic follows a two-step waterfall execution: first, it queries the local cache for an active instance (Hit); if unavailable, it scans for neighbors (Partial Hit); finally, if both fail, it defaults to the cloud fallback (Miss).
    \end{enumerate}

    \begin{figure}[htbp]
\centering
\includegraphics[width=1\columnwidth]{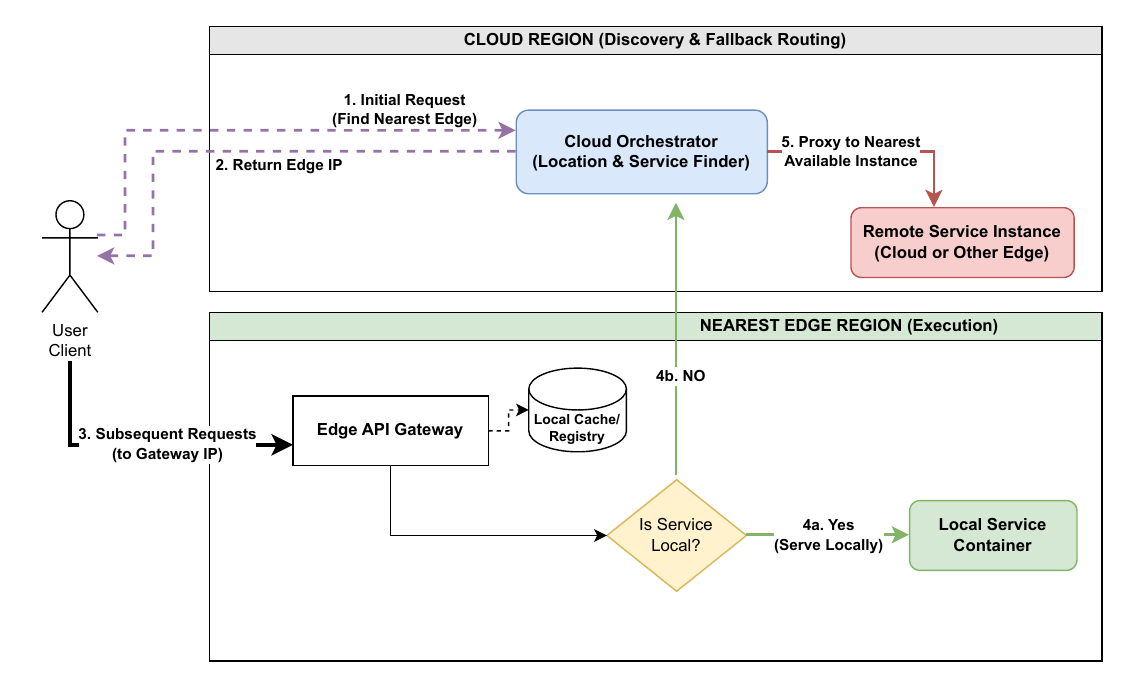}
\caption{Operational Workflow of the Request Routing in proposed Dynamic Cloud–Edge Architecture}
\label{edge}
\end{figure}

Figure \ref{edge} depicted the operational workflow of request routing in the proposed dynamic cloud-edge architecture. The cloud orchestrator responds to an initial client request to determine the closest edge gateway, then uses cached service information to route subsequent requests directly to the edge. When possible, incoming requests are handled locally at the edge; otherwise, the cloud fallback ensures service continuity during cold starts or periods of unavailability.

\item \textbf{Telemetry and Asynchronous Feedback Loop:} To ensure the Orchestrator makes informed scaling decisions without introducing latency into the critical data path, the framework employs a decoupled monitoring strategy. This asynchronous feedback loop consists of two coordinated phases:

    \begin{itemize}
        \item Metrics Aggregation: First, during metric instrumentation and buffering, the edge gateway used the micrometer to collect real-time metrics, such as http\_server\_requests and active\_user\_sessions. Crucially, these metrics are not streamed in real time; instead, they are buffered in a sliding time window to decouple telemetry overhead from request processing.
        \item Periodic State Synchronization: To transmit the data, a background task is executed at a fixed interval (configured at 30 seconds), i.e., (@Scheduled(fixedDelay = 30000)) \footnote{\textit{The source code and implementation details are available at the} \href {https://bitbucket.org/hulkhire/edge_project/src/main/}{Bitbucket Repository}}. It aggregates the buffered metrics into a lightweight JSON payload. That payload is transmitted to the cloud orchestrator via a secure REST endpoint, ensuring the control plane receives consistent state updates without blocking the data plane.
    \end{itemize}

\item \textbf{The Orchestrator and Hysteresis Logic (Control Plane):}
The Orchestrator serves as the decision engine responsible to translate the aggregated telemetry data into actionable deployment commands. To balance the system responsiveness with operational stability, the control logic implements a two-stage evaluation process:
    
\begin{itemize}
    \item{Traffic Analysis and Stabilization:} To prevent rapid oscillation between instantiation and termination of the microservice  the Orchestrator utilizes a 1-minute Simple Moving Average (SMA) on incoming logs. It validates traffic against the following hysteresis thresholds:
    
    \begin{itemize} 
        \item $T_{high}$ (Activation Threshold): Set to 80 concurrent users.
        \item $T_{low}$ (Deactivation Threshold): Set to 30 concurrent users.
    \end{itemize}
    
    \item{Dynamic Lifecycle Execution:} Based on the hysteresis evaluation, the Orchestrator issues state-change commands via the Docker Remote API:

    \begin{itemize}
        \item Scale Up Event: If $SMA > T_{high}$ and there is no local container, the Orchestrator sends a command to the appropriate Edge Node's Docker Daemon to launch the microservice image.

        \item Scale Down Event:  If $SMA < T_{low}$ and a container is running, the Orchestrator sends a docker stop command to the edge node, which puts it back in its idle, energy-saving state.
    \end{itemize}
\end{itemize}

Threshold values are selected empirically based on preliminary load characterization and can be tuned dynamically without altering the orchestration logic.

In the experimental setup, the above Activation threshold $(T_{high})$ and deactivation $(T_{low})$ threshold values are intentionally increased in certain scenarios to effectively disable edge activation and replicate a cloud-only execution environment. It enabled a fair, controlled comparison of cloud-only, hybrid, and gateway-based distributed deployments without modifying the underlying orchestration logic.
      
\item \textbf{Service Convergence and State Synchronization:} A critical challenge with dynamic orchestration is that there is a time difference between container instantiation and service readiness. And we utilize the Eureka self-registration method to create an eventually consistent model so that traffic only hits properly initialized instances.
    
    \begin{itemize}
        \item Startup: When the Edge Container starts, it automatically sends a heartbeat to the Cloud Eureka server.

        \item Discovery: The Cloud Eureka updates its registry.

        \item Propagation: During the next 30-second fetch cycle, the Edge Gateway downloads the new registry, sees the local instance, and automatically switches traffic from the Cloud Fallback to the Edge Container.
    \end{itemize}
\end{enumerate}

\section{Performance Analysis}
The proposed hybrid cloud–edge orchestration model is evaluated across a range of deployment scenarios to assess its impact on the efficacy of cloud offloading, execution stability, and end-to-end latency. The performance analysis considers system behaviour during scaling transitions, as well as tail and average latency. These factors are particularly critical for latency-sensitive applications and services that must adhere to stringent Service Level Agreement (SLA) requirements. The experimental results show that the model achieves a latency reduction of up to 16.37\% compared to the traditional cloud-only model. Also, the analysis of request logs shows a persistent improvement in response time when requests are handled by the edge layer, primarily due to the elimination of wide-area network propagation delays and the efficiency of gateway-level service discovery.

To better understand the behavior of the system at multiple scales, three deployment scenarios are evaluated:
\begin{itemize}
    \item \textbf{Single Edge Scenario:} An initial setup in which there is just one active edge node.
    \item \textbf{Double Edge Scenario:} To test load distribution, an intermediate setup where two edge nodes are operational.
    \item \textbf{Triple Edge Scenario:} To determine saturation limits, a maximum load setup is tested where all three edge nodes are operational.
\end{itemize}
Detailed logs are collected for each scenario, documenting request delays, the execution location (cloud or edge server), and request responses.

\subsection{Experimental Setup and System Configuration:} The cloud orchestrator and fallback microservices are deployed on Amazon Web Services (AWS) EC2 instances, equipped with one CPU core and around 1 GB of memory, operating on Ubuntu 22.04 LTS. To ensure scalability and fault tolerance, multiple instances can be deployed across different availability zones, with load balancing mechanisms in place to handle increased traffic and redundancy.

The edge layer uses several small-form-factor computing devices poared by Intel Core i7 (x86\_64) processors with multi-core architectures and hyper-threading. Each edge node is equipped with 8 GB of DDR4 RAM and 1 TB of SSD storage, representing actual resource-limited edge deployment scenarios. All edge devices ran Ubuntu 22.04 LTS (64-bit) and used Docker for containerization, with resource isolation implemented via container-level control mechanisms. This detailed hardware overview is intended to give the audience confidence in the system's performance capabilities.

Also, the communication between cloud and edge components is facilitated by SSH tunnelling alongside the Remote Docker API, enabling safe control and management of distributed edge containers from the cloud. This approach is designed to foster trust in the system's secure and reliable communication channels. 

\subsection{Latency Distribution and Scalability Analysis}

When only one edge server is running, the average latency drop achieved by the proposed model is approximately 14.47\%. Moreover, edge-based execution not only reduced average response time but also improved performance stability. The requests processed by Cloud showed considerable latency variation, with some taking significantly longer due to internet routing. However, the requests processed on the edge are completed within a much narrower time frame, indicating more predictable and reliable response times. The reduction in latency variance highlights the effectiveness of localized execution in improving response-time determinism.

\begin{figure}[htbp]
    \centering
    \includegraphics[width=\linewidth]{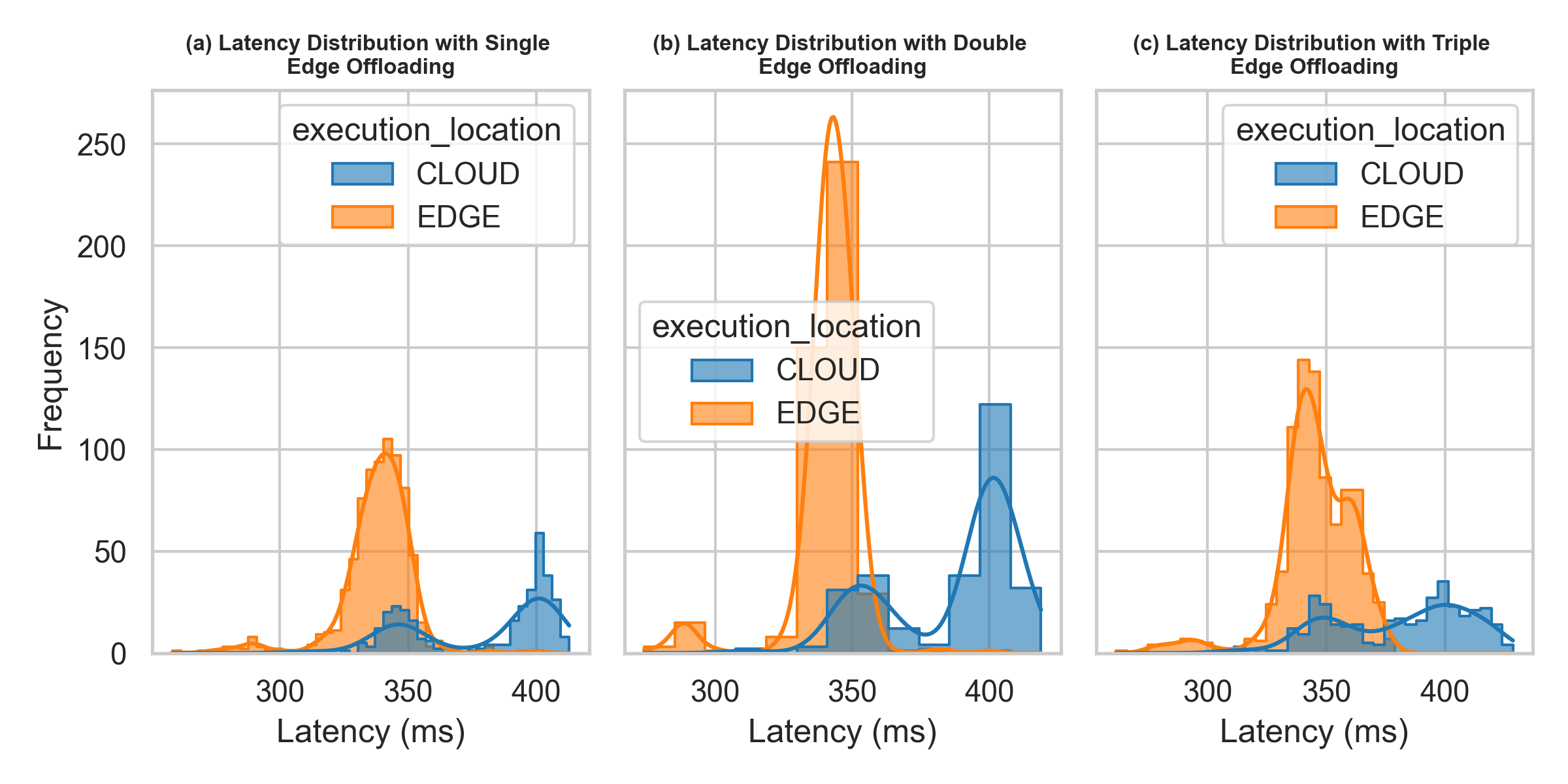}
    \caption{Impact of Edge Node Scaling on Cloud and Edge Latency Distributions. Latency distributions are shown for (a) single, (b) double, and (c) triple-edge server configurations.}
    \label{fig:edge_latency_scaling}
\end{figure}
Figure \ref{fig:edge_latency_scaling} shows how the latency distribution of cloud and edge execution varies with different edge deployment scales. (a) With just one edge server, edge requests already have lower and more stable latency than cloud execution. Increasing the number of edge servers, as shown in (b) and (c), narrows the edge latency distribution and shifts it further left, signifying enhanced load distribution and reduced response times compared to the cloud.

\subsection{Edge Cluster Scaling Analysis}
Further experiments are conducted by increasing the number of active edge servers. When two edge servers are enabled, the proposed system achieved the best performance, with an average latency reduction of 16.37\%. The latency drop is achieved because the orchestrator can distribute incoming requests more effectively, without overloading any edge servers. However, when three edge servers are enabled concurrently, the latency reduction is mitigated to about 12.87\%. And the nonlinear scaling does not reflect algorithmic limitations in the orchestration logic but is rather an artifact of the experimental testbed, where multiple edge instances are simulated using nested containerization on a single physical host. As more active containers are used, CPU scheduling contention and hypervisor overhead start to offset the benefits of edge proximity. Such resource contention would not be present in a physically distributed industrial deployment, where linear scalability can be expected.


\begin{figure}[htbp]
    \centering
    \includegraphics[width=\linewidth]{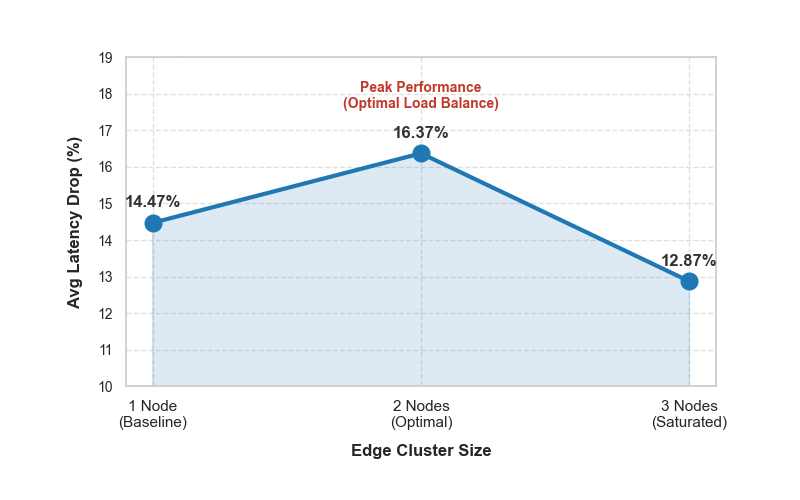}
    \caption{Impact of Edge Cluster Size on Average Latency Reduction.}
    \label{fig:edge_cluster_vs_latency}
\end{figure}

Figure \ref{fig:edge_cluster_vs_latency} deliberates how latency drops oscillate with the variable number of active edge nodes. The proposed model achieves peak efficiency with a dual-node configuration at 16.37\%, but, due to hardware resource contention and saturation in the triple-node scenario, it mitigates the net gain to 12.87\%.

The given observation is further strengthened by the comparison of tail latency as shown in Figure \ref{fig:tail_latency}, where the comparison of average, P95, and P99 latencies is carried out for single, double, and triple-edge configurations. Meanwhile, the double-edge configuration achieved the lowest average latency, with the P95 and P99 latency metrics exhibiting diminishing returns and, in some cases, slight regression for the triple-edge configuration. The difference between the average and tail latencies suggests a possible increase in latency variance due to CPU scheduling contention and container runtime overhead when multiple edge instances are packed together on the same physical machine. Significantly, the tail latency degradation is not a consequence of the orchestration logic but is an artifact of the experimental setup, where multiple edge containers are competing for limited shared compute resources. In a physical edge deployment setup, such conflict is expected to be alleviated, and average and tail latencies are expected to improve consistently as edge density increases.

\subsection{Tail Latency Analysis (Average, P95, and P99)}
While average latency provides a general system performance indication, sometimes it often conceals severe delays that can have a critical effect on the user experience. To understand such a worst-case scenario, percentile-based latency metrics such as 95th percentile latency (P95) and 99th percentile latency (P99) are commonly used in latency-sensitive systems. P95 is the latency threshold for the top 95\% of requests, and P99 indicates the tail distribution of the worst 1\% of requests. These metrics are very helpful for measuring Service Level Agreement (SLA) compliance, which is typically violated by tail events rather than the mean performance. The latency data is analyzed to find the average, P95, and P99 latency values for all deployment scenarios and compared with the single-edge baseline. The experimental results demonstrate that moderate edge scaling, especially in the dual-edge configuration, results in more stable tail latency by limiting extreme latency outliers, while augmenting the edge nodes brings only a marginal change in the variation of P95 and P99 values due to shared resource contention in the emulated environment. These findings validate that the proposed orchestration framework enhances worst-case latency stability and reveal the sensitivity of tail latency to resource isolation at higher edges.

\begin{figure}[htbp]
    \centering
    \includegraphics[width=\linewidth]{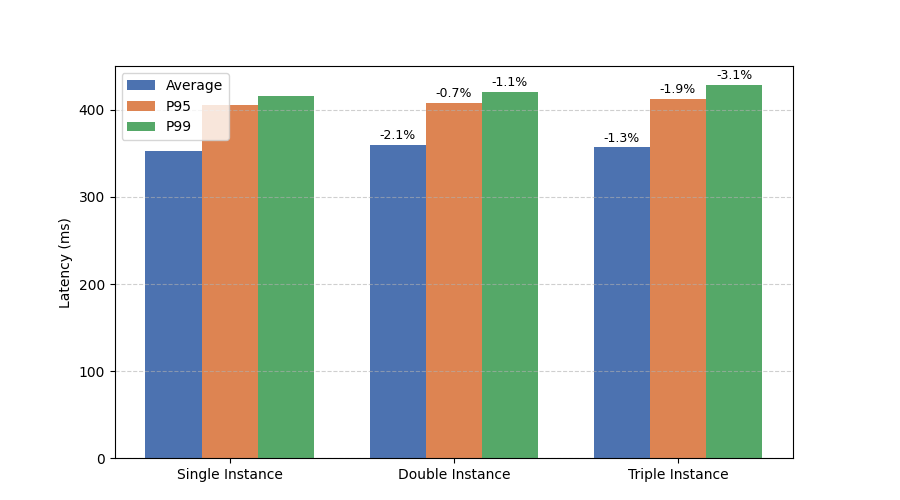}
    \caption{Tail Latency Comparison (Average, P95, and P99) Across Edge Deployment Scenarios}
    \label{fig:tail_latency}
\end{figure}

Figure \ref{fig:tail_latency} displays a comparative analysis of average, P95, and P99 latency values for single-edge, double-edge, and triple-edge deployment scenarios. Although modest gains in average latency have been observed, tail latency values reveal how the system performs under peak loads and bursty conditions. 
The dual-edge deployment configuration shows the most stable tail performance, while the triple-edge deployment topology shows only a slight variation in P95 and P99 latency due to resource contention in the emulated environment. The above results indicate that the proposed orchestration framework is able to stabilize the worst-case latency values when there is moderate edge scaling, while the tail latency values become more sensitive to resource isolation with increased edge density. The analysis emphasizes the importance of considering both average and tail latency while evaluating edge orchestration frameworks for latency-sensitive applications. Negative values indicate latency regression due to CPU scheduling contention in the emulated saturation scenario.

\begin{figure}[htbp]
    \centering
    \includegraphics[width=\linewidth]{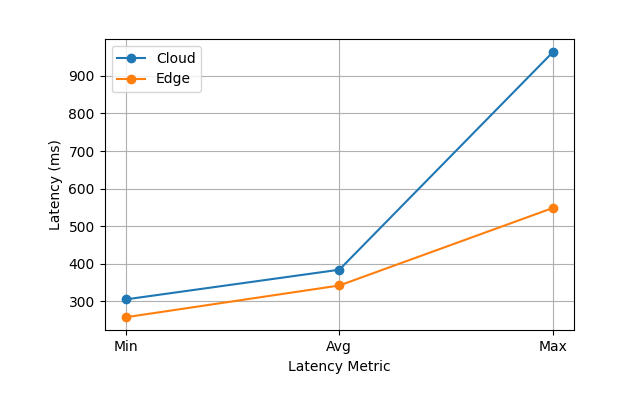}
    \caption{Minimum, Average, and Maximum Latency Comparison for Cloud Fallback and Edge Execution}
    \label{fig:min_max_avg}
\end{figure}

Figure \ref{fig:min_max_avg} reflects the comparison of minimum, average, and maximum latency observed for requests served by the cloud fallback and the edge within the proposed hybrid framework. Edge execution consistently exhibits lower latency across all metrics, with a notably reduced maximum latency, indicating improved worst-case performance compared to cloud fallback.

\subsection{Cloud Offloading Efficiency}

Apart from the latency improvements, the proposed framework also demonstrates a substantial reduction in cloud dependency. The aggregated request logs for all three scenarios, i.e., single node, dual-node, and triple-node, indicate that roughly two-thirds of the total requests are being satisfied by the edge nodes, while the rest are handled by the cloud fallback system. During the initial service cold-start phases or in exceptional conditions, the central cloud primarily serves as a fallback system. The behavior contributes directly to reduced backhaul bandwidth consumption and improved scalability.\\

\begin{figure}[htbp]
    \centering
    \includegraphics[width=0.8\linewidth]{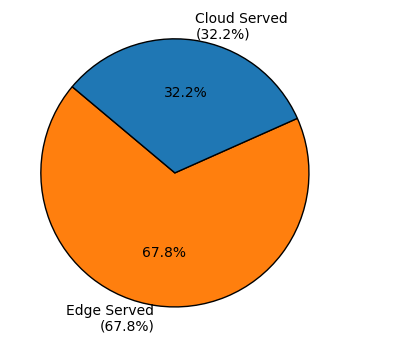}
    \caption{Distribution of Requests Served by Edge and Cloud}
    \label{fig:edge_request_distribution}
\end{figure}

Figure \ref{fig:edge_request_distribution} displays the ratio of client requests handled by edge servers versus the server present at the cloud layer. Approximately two-thirds of requests, i.e., 67.8\%, are served at the edge, significantly reducing the load on the cloud and minimizing wide-area network communication. The behavior helps in contributing directly to reduced backhaul bandwidth consumption and thus in improving scalability.

\subsection{Cold-Start and Transition Behavior}
Dynamic edge orchestration introduces a cold-start period where the services initially are not available at the edge, and the requests are processed by the cloud. To analyze this behavior, a timeline of the cold-start latency is extracted from the request logs to understand the transition from cloud-based execution to stable edge execution. 

The results indicate a sustained reduction in latency immediately after edge activation.  After the edge service becomes available, the request latency stabilizes at a lower value without any fluctuations, which confirms that the hysteresis-based orchestration strategy involves frequent scale-in and scale-out events while ensuring smooth transitions.

\begin{figure}[htbp]
    \centering
    \includegraphics[width=\linewidth]{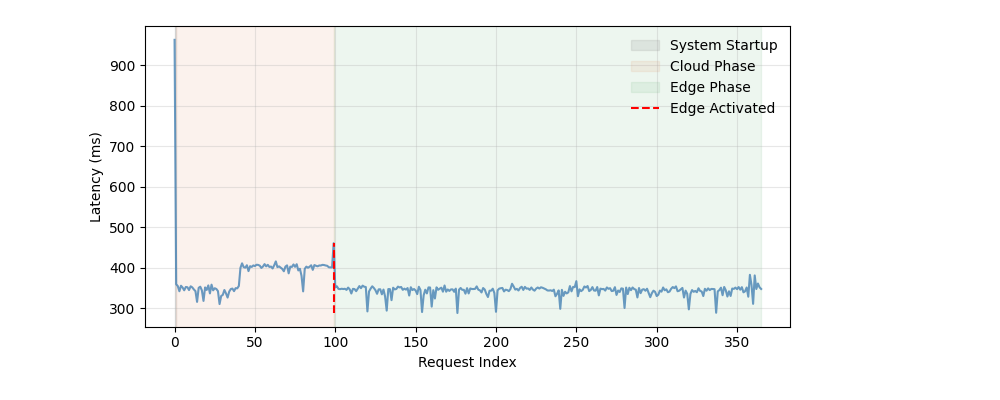}
    \caption{Cold-Start Latency Evolution with Cloud-to-Edge Execution Phases}
    \label{fig:cold_start}
\end{figure}

The cold-start behavior of the proposed hybrid cloud-edge system is shown in Figure \ref{fig:cold_start}, which plots the request-level latency across the execution process. During the system startup and initial cloud-only phase, the latency of the requests is higher and more volatile because of the wide area network traversal and absence of active edge services. The transient latency peak noticed around the point of edge activation represents the initial request processed during the container initialization and service registration, where routing overlaps with environment startup latency. Once the edge instance is fully available, a clear and sustained latency reduction is noticed, followed by a stable execution phase with negligible latency variation. The coherent behavior clearly indicates that the proposed orchestration framework enables a smooth cloud-to-edge transition while avoiding any oscillatory behavior, reflecting predictable performance after edge activation.

The same behavior is also further reflected in the aggregate latency statistics, where the maximum latency values for cloud execution are significantly higher than those observed at the edge, as shown in Figure \ref{fig:min_max_avg}, confirming that cold-start and fallback activities are the primary contributors to worst-case cloud latency

It is important to note that the framework aims to maintain service continuity rather than to mitigate container startup time. Although it is important to aim to minimize initialization delay, our architecture ensures that zero requests are ever dropped throughout the cold start phase by routing them to Cloud Fallback smoothly.  The latency disparity, as revealed in Figure \ref{fig:cold_start}, quantifies the \quotes{cost} of this fallback, which is paid in favour of ensuring 100\% availability over instant edge execution during the transition.

\subsection{Comparative Evaluation}
To assess the effectiveness of the proposed dynamic orchestration framework, experiments are carried out under three execution scenarios: 
\begin{itemize}
    \item a cloud-only deployment, where all requests is handled centrally,
    \item a hybrid cloud-edge deployment with distributed edge gateways, and
    \item  a hybrid deployment with a centralized gateway hosted in the cloud.
    
\end{itemize}

In the cloud-only configuration, all incoming requests are handled by the central cloud service (i.e., the fallback service), resulting in higher end-to-end latency due to traversing wide-area networks. In the cloud-only deployment, edge deployment is disabled by threshold tuning, ensuring that all requests are handled by the cloud. However, in the distributed hybrid deployment, the best results are obtained, with latency dropping by up to 16.37\% as requests are dynamically routed to geographically close edge nodes after services are instantiated. In the centralized gateway deployment, requests are handled by a single gateway instance that is hosted in the cloud. Although the orchestration logic remained active, all requests continued to be handled by the cloud-hosted fallback service, and there is no notable migration of edge services. As a result, the system largely relied on the cloud-hosted fallback microservice to handle requests. Such behavior shows that the centralized gateway's high dependency on cloud resources makes the edge infrastructure ineffective, which contradicts the main goal of autonomous edge offloading.

In order to contextualize the effectiveness of the proposed framework, we compare it with the recent state-of-the-art orchestration techniques published between 2024 and 2025. Table \ref{tab:comparison2} provides the comparative study of the proposed framework against recent developments in Deep Reinforcement Learning (DRL), Large Language Models (LLMs), and bio-inspired optimization models. The impact of horizontal line analysis emphasizes the trade-off between latency reduction capability and computational complexity, which evaluates whether the structural optimization at the architectural level can provide a more feasible alternative to algorithmic tuning in resource-constrained edge environments.

\begin{table*}[t]  
\begin{adjustwidth}{-1cm}{-1cm}
\centering
\scriptsize
\setlength{\tabcolsep}{4pt}
\renewcommand{\arraystretch}{1.3}
\newcolumntype{Y}{>{\centering\arraybackslash}X}
\begin{tabularx}{1\textwidth}{|p{2cm}|Y|Y|Y|Y|Y|Y|Y|}
\hline
\multirow{2}{*}{\textbf{Schemes}}
& \multicolumn{4}{c|}{\textbf{Comparison parameters}} \\
\cline{2-5}
& \textbf{Core Methodology}
& \textbf{Orchestration Goal}
& \textbf{Performance Metrics}
& \textbf{Operational Overhead} \\
\hline
Peng et al. \cite{AIservices}
& AMDQN (Deep RL)
& Service Chaining
& 10 \% (vs Adv. DRL)/ 41\% (vs. Greedy)
& High (Training + Convergence) \\
\hline

Feng et al. \cite{GenAI}
& Large Language Models (LLM)
& Deployment Decisions
& 22.8\% (During Surges)
& Extremely High (Inference Latency) \\
\hline

Damera et al. \cite{damera2024snake}
& Improved Snake Opt. (ISO)
& Task Scheduling 
& 6\% Makespan Improvement
& Moderate (Stochastic Iterations) \\
\hline

Mahapatra et al. \cite{EnergyAware}
& Fuzzy Logic + BLW-JAYA
& Energy \& Load Balancing
& 7\% Latency Improvement
& Low–Moderate (Rule Evaluation) \\
\hline
Proposed Work
& Decoupled Gateway + DBSCAN
& End-to-End Latency
& 6.37\% (vs.Cloud Baseline)
& Negligible (Deterministic) \\
\hline

\end{tabularx}

\caption{Comparative Analysis with State-of-the-Art Orchestration Frameworks.}
\label{tab:comparison2}
\end{adjustwidth}
\end{table*}







Table \ref{tab:comparison2} highlights an important diminishing return effect within recent advancements in algorithmic optimization methods. While advanced learning-based models like AMDQN \cite{AIservices} and LLM-driven orchestration \cite{GenAI} are found to achieve high absolute gains against naive baselines, like Greedy (with improvements of up to 41\%), their marginal improvement over existing advanced algorithms is often limited (with improvements of up to 10.13\%). Furthermore, these gains are accompanied by substantial operational overhead, including GPU-based training cycles and non-deterministic inference latency, which limits their practicality on resource-constrained edge gateways.

In contrast, the proposed framework achieves the strong end-to-end latency drop of 16.37\% strictly through architectural decoupling. By removing synchronous wide-area control-plane interactions from the critical request path via distributed gateways, the framework delivers deterministic performance improvements without incurring the computational overhead associated with training or executing learning-based models. These results propose that structural optimization offers a highly scalable and predictable alternative to complex algorithmic tuning for real-time edge orchestration.


\begin{figure}[htbp]
    \centering
    \includegraphics[width=\linewidth]{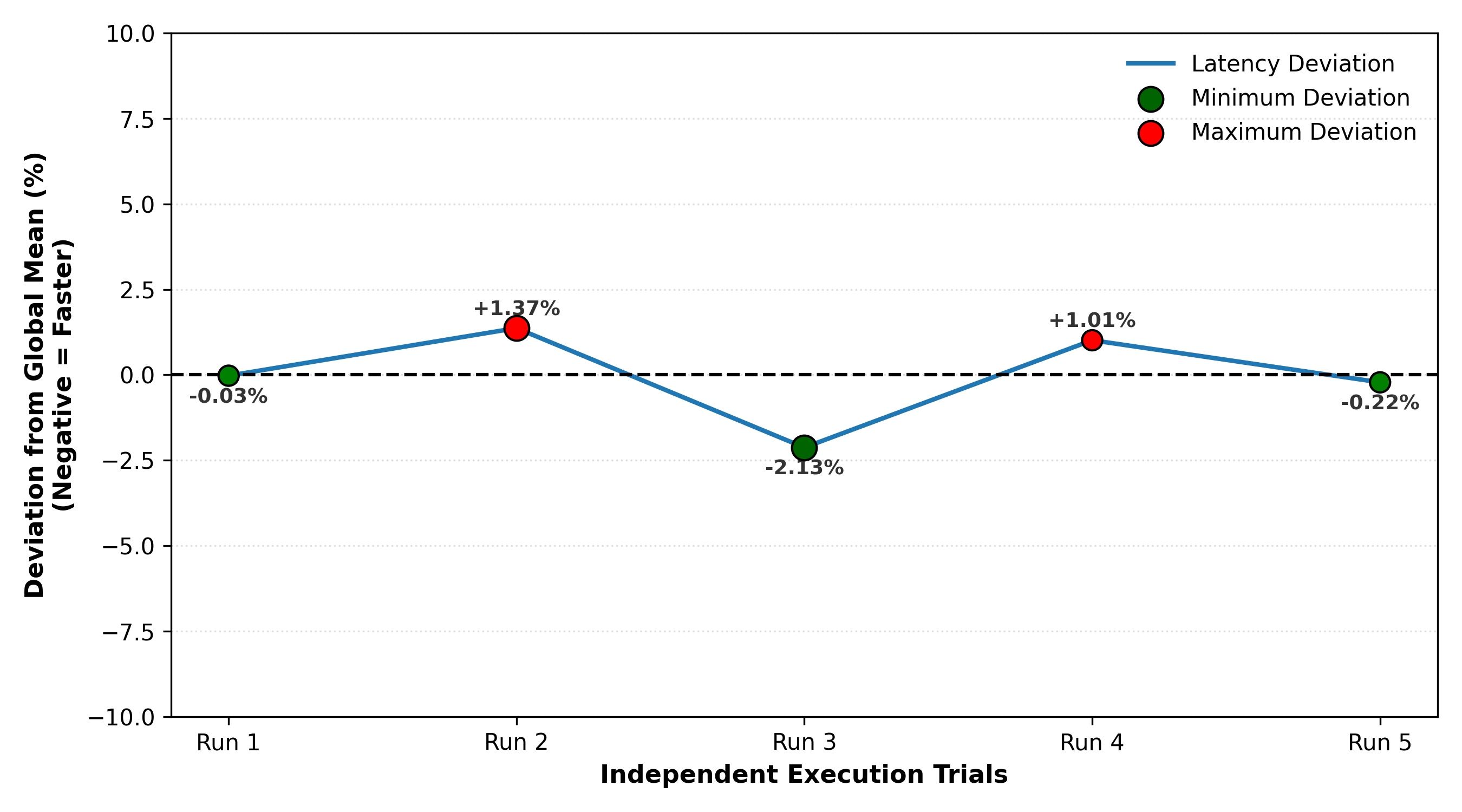}
    \caption{Latency Deviation Across Repeated Experimental Executions}
    \label{fig:deviation}
\end{figure}

Figure \ref{fig:deviation} displays the percentage deviation of average request latency across repeated experimental executions relative to the global mean value. The range of deviation for the experiment is strictly included in the interval from -2.13\% to +1.37\%, indicating the highly stable and deterministic behavior of the framework under test. The minimum modulus deviation value is -0.03\%, representing near alignment to the global average, while the maximum modulus deviation value defines the worst-case variability boundary at -2.13\%. Such minor percentage deviation rates can be attributed due to the scheduling overhead of the containers, the transient SSH-based testbed networking jitter, the processing latency of cloud services and sometimes applicable public network bandwidth rather than the orchestration solution. Furthermore, the fact that all the rates are below the range of ±2.13\% ensures that the proposed framework effectively absorbs system- and network-level variability, delivering predictable latency performance suitable for industrial SLA requirements.

In summary, the above results confirm that the proposed model not only optimizes the response time but also achieves more stable performance and efficient use of network resources, making it more suitable for latency-sensitive and bandwidth-constrained networks.

\section{ Discussion and Industrial Implications}

The proposed work demonstrates a practical implementation of a dynamic cloud-edge orchestration framework focused on mitigating latency for geographically distributed, latency-sensitive applications \cite {liu2023realtime,latency_aware}. The system consists of a centralized cloud orchestrator, multiple distributed edge gateways, and dynamically deployed microservices. By using location-based request routing, service discovery caching, and demand-based service placement, computation is moved closer to end users whenever possible.

In terms of response time, hybrid cloud-edge execution consistently outperforms cloud-only execution, a primary finding of the experimental evaluation of the proposed work. The focus on stability through cached service discovery at edge gateways can reassure IT professionals and researchers about the system's robustness. Reduced Wide Area Network (WAN) traversal and the hybrid architecture's ability to route requests locally without sending them to the cloud registry on every request are the primary factors contributing to the latency improvements. In this instance, the cached service discovery at edge gateways decoupled the data plane from the control plane, thereby enhancing the system's stability even under variable network conditions. This behavior is further supported by the run-to-run latency variability, which is within 2.13\%, thereby ensuring deterministic performance across repeated experimental executions. 

The experimental results also emphasize how the edge density affects the performance. While increasing the number of active edge servers reduces latency by more evenly distributing the request load, the associated overhead suggests that future work should explore scalable deployment strategies. This behavior is not a limitation of the proposed logic itself but rather an implication of the deployment environment. In the proposed architecture, edge servers are implemented on a single physical machine using containerization with Docker, and these servers are exposed to the cloud through SSH reverse tunneling. As the number of edge servers increases, CPU scheduling contention and container management overhead begin to offset the benefits of reduced network distance.

\subsection{Limitations}

It is essential to evaluate the validity of the experimental testbed. Although the assessment is based on containerized emulations on a single physical host, the achieved latency drop of 16.37 \% are supported by SSH tunneling. This approach maintains protocol interactions between cloud and edge components, allowing researchers to confidently compare orchestration behaviors. By isolating orchestration overheads and routing logic from uncontrolled physical network variability, the experiment underscores the impact of decoupling the control plane from the data plane on request latency. Since the architecture eliminates synchronous control plane interactions from the critical request path, the relative performance improvements over centralized baselines remain significant. While SSH tunneling does not fully capture the variability found in wide-area networks, it preserves the protocol-level interaction between cloud and edge components, facilitating meaningful comparisons of orchestration behaviors.

In addition, the centralised gateway configuration imposes certain constraints on the proposed model. The cloud host co-locates the failover microservice and the gateway in the provided configuration. Consequently, the gateway inherently prioritizes locally available cloud services and transfers requests to the cloud, even when peripheral resources are logically eligible for execution. Because the centralized gateway is designed to prioritize local cloud services, the movement of services and the utilization of peripheral resources are restricted by the underlying behavior, rather than a deficiency in the orchestration logic. Recognizing these limitations can motivate professionals to develop more flexible architectures that better leverage edge proximity, which is crucial for advancing edge computing capabilities. In addition, to facilitate efficient edge utilization in a centralized gateway architecture, either maintain distinct gateway instances or implement supplementary coordination mechanisms across multiple cloud servers. Recognizing these limitations can inspire researchers and engineers to explore innovative solutions, fostering a sense of opportunity for future advancements in offloading strategies. 

Although the above limitations are present, the proposed framework demonstrates its core objective: offloading a significant portion of client requests from the cloud server to the edge server. Based on the steady-state analysis, it is identified that the majority of requests are being served locally. These characteristics make the framework suitable for real-world scenarios such as IoT data ingestion, smart city platforms, and mobile sensing applications, which can inspire further exploration of its potential benefits.

\subsection{Industrial implications}
The proposed dynamic cloud-edge orchestration paradigm offers a clear advantage in reducing costs, helping organizations feel confident about financial efficiency. It enables a transition from rigid, capital-intensive infrastructure to an agile, OpEx-optimized deployment model, which can alleviate concerns about high infrastructure expenses. By dynamically instantiating edge microservices only when demand exceeds thresholds, organizations can better manage resources and costs during peak loads.  Critically, SLA compliance depends on the framework’s ability to stabilize latency during dynamic workloads, ensuring reliable performance for decision-makers. It provides the deterministic latency needed for smart city and industrial IoT applications, contrasting with cloud-only setups where network variability can cause issues. Additionally, maintaining around 67.8 \% of traffic locally cuts cloud egress costs and backhaul bandwidth, supporting operational efficiency.

Operational managers need to understand the limitations of hardware. Scaling is only effective if the physical servers can handle the software load. Our experiment demonstrated that placing too many containers on limited hardware, as we did in the triple-edge server scenario, can actually decrease performance. Therefore, a successful deployment requires investing in adequate physical processing power to meet the software's demands, rather than relying solely on intelligent software logic or complex fuzzy-based resource management \cite{chen2021fuzzy, jin2024adaptive}.

\section{Conclusion and Future Work}

The proposed work introduces a dynamic cloud-edge orchestration framework designed to reduce end-to-end latency in geographically distributed applications. By explicitly separating the control plane from the data plane and adopting a lightweight, demand-aware deployment approach, the architecture addresses key routing bottlenecks common in centralized cloud setups. A detailed experimental evaluations demonstrate that the proposed model can achieve a latency reduction of up to 16.37\% compared to cloud-only execution. It improves response time stability, including tail latency behavior (P99), while effectively offloading backhaul traffic. Notably, the framework shows that decentralised gateway intelligence can deliver consistent performance improvements without the high computational complexity of advanced learning-based solutions. So, this makes it particularly suited for edge environments with limited resources where ease of use is critical. 
Despite the fact that the comprehensive experimental analysis is performed on a containerized emulation testbed, the saturation behavior suggests that the scalability is robust and could be further enhanced in physically distributed environments with dedicated network connectivity. Overall, the proposed architecture provides a flexible, practical foundation for developing next-generation cloud-edge systems that are fast, reliable, and efficient. 

Moreover, the future work will focus on testing the solution across various edge hardware configurations and on exploring predictive orchestration based on workload forecasting to proactively handle demand peaks.

\bibliographystyle{elsarticle-num} 
\bibliography{refrence}
\end{document}